\documentclass[a4paper]{spie}  %>>> use this instead for A4 paper
\usepackage[]{graphicx}
\usepackage[utf8]{inputenc}
\usepackage[symbol]{footmisc}

\usepackage{indentfirst}

\title{Cosmology with SKA\footnote[2]{\quad Based on a talk delivered by one of us (O.B.) at the SKA.PT Days at February 7th, 2018 in Lisbon, Portugal} }

\author{Orfeu Bertolami and Cláudio Gomes%\supit{q}
\skiplinehalf
Departamento de Física e Astronomia, Faculdade de Ciências da Universidade do Porto, Rua do Campo Alegre s/n 4169-007 Porto, Portugal.
}

\authorinfo{E-mail: orfeu.bertolami@fc.up.pt}
\begin{document} 
  \maketitle 

%%%%%%%%%%%%%%%%%%%%%%%%%%%%%%%%%%%%%%%%%%%%%%%%%%%%%%%%%%%%% 
\begin{abstract}
We review some of the major contributions that the Square Kilometer Array (SKA) will provide for Cosmology. We discuss the SKA measurements of the equation of state parameter for dark energy from Baryonic Acoustic Oscillations (BAO), of the dark matter power spectrum and modifications of the Poisson equation or the slip relation from weak lensing. We also comment on measurements of the cosmic magnetism and its role on the dynamics of the Universe.
\end{abstract}

%>>>> Include a list of keywords after the abstract 


%%%%%%%%%%%%%%%%%%%%%%%%%%%%%%%%%%%%%%%%%%%%%%%%%%%%%%%%%%%%%
\section{INTRODUCTION}
\label{sec:intro}  % \label{} allows reference to this section

The Square Kilometer Array (SKA) is a worldwide collaboration with radiotelecopes hosted in South Africa and Australia at a first stage, and in other eight African countries later on, namely, Botswana, Ghana, Kenya, Madagascar, Mauritius, Mozambique, Namibia and Zambia. This will be largest radiotelescope array in the world and it will end up being 50 times more sensitive and 100 times faster than current best radiotelescopes. Its life span is expected to be of at least 50 years, and although it uses the most recent antenna technology and signal processing and computing, it can be continuously upgraded as computing power increases. SKA will cover frequencies in the range from 70 MHz to 10 GHz.

	The knowledge of the Universe will be dramatically increased with SKA, since it will provide insight on the evolution of galaxies and cosmology, on strong gravity through pulsars and black holes, on the origin and evolution of cosmic magnetism, on cosmic history of the Universe at dark ages and reionisation epochs, and on putative cradle of life\cite{ska}.
	
	In what concerns Cosmology, radio observations from SKA are expected to constrain the equation of state parameter for dark energy, hence discriminating between several models of cosmic acceleration, to observe a dark matter power spectrum, to test modified theories of gravity using weak lensing observations, and to get and insight on the cosmic magnetism\cite{ska,bull,raccanelli}. 

%%%%%%%%%%%%%%%%%%%%%%%%%%%%%%%%%%%%%%%%%%%%%%%%%%%%%%%%%%%%%
\section{COSMOLOGY} 

\subsection{Dark Universe} 

One of the science key goals of SKA concerns the understanding of the dark components of the Universe. Several missions have been designed to study the energy content of the Universe. The most recent, Planck mission\cite{planck}, showed that the Universe is filled with $4.9\%$ of ordinary matter, $26.8 \%$ of dark matter and $68.3\%$ of dark energy. Dark matter is responsible for the flattening of the galactic rotation curves, and gravitational lensing which cannot be explained only by regular matter. On its turn, dark energy is the smooth energy substratum behind the current acceleration of the Universe. These two dark components dominate the Universe. Nevertheless, aside from some observational signatures and dynamical effects, their inner nature still remains unknown.

\begin{figure}
\centering
\includegraphics[width=0.7\textwidth]{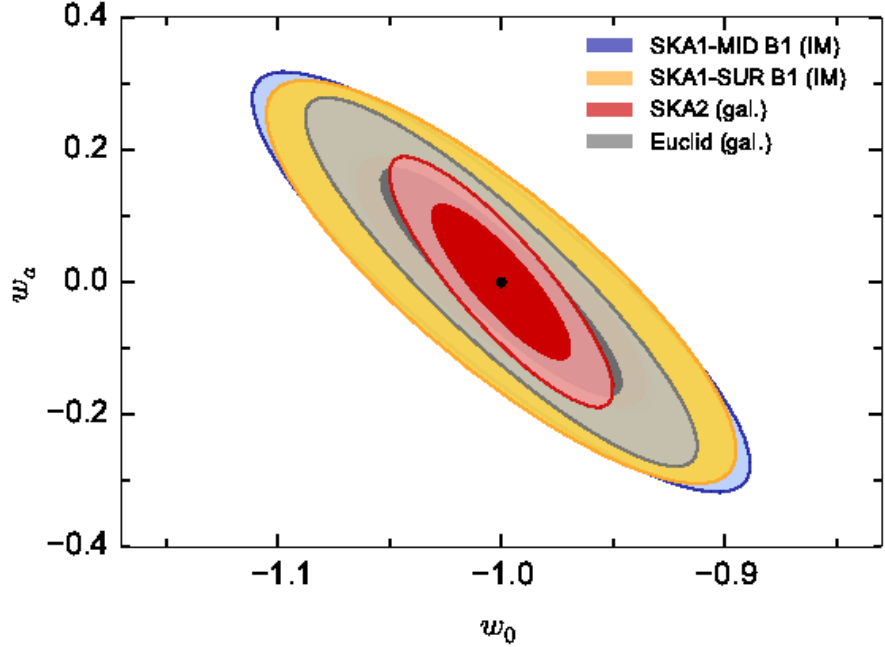}
\caption{Predicted SKA constraints at various stages on the equation of state parameter of dark energy (with BOSS and Planck priors) \cite{raccanelli}.}
\label{fig:de}
\end{figure}

	SKA will be able to provide some answers about the components of the Universe through the 21cm radiation of the neutral hydrogen at high redshifts. The cosmic evolution of such well defined signal encodes information about the medium where it propagates, most particularly about the dark matter and dark energy effects. For instance, by measuring the BAO signature in the radio of the galaxy power spectrum, SKA will allow for precise measurements of the equation of state of dark energy, $w=p/\rho$, of the form:
	\begin{equation}
	w=w_0+w_a(1-a)~,
	\label{eq:1}
	\end{equation}
where $p$ and $\rho$ are the pressure and energy density of dark energy, $w_0$ and $w_a$ are some real parameters and $a$ is the scale factor. In Fig. \ref{fig:de} one can see that assuming a Gaussian distribution around the fiducial model $w=-1$, which corresponds to the cosmological constant as dark energy, the predicted SKA results combined with Planck \cite{planck} and Baryon Oscillation Spectroscopic Survey (BOSS) \cite{BOSS} priors constrain considerably the allowed region of parameters. The expected combined precision is compared with the one from Euclid mission \cite{euclid}.
%%%%%Sometimes it is necessary to precede the double slash 
%%%%%by \verb|\protect| to get the desired result, 
%%%%%for example, in article titles.

Furthermore, by studying weak gravitational lensing the measurement of the dark matter power spectrum will be possible, and bounds on the mass and the number of neutrinos families can be obtained.

\subsection{Testing gravity to the limit}

\begin{figure}
\centering
\includegraphics[width=0.7\textwidth]{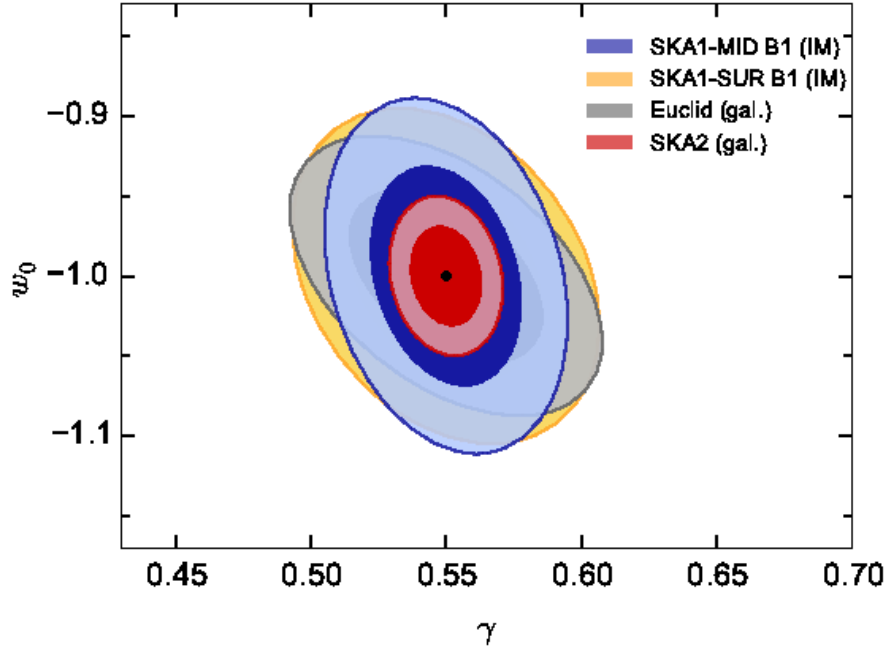}
\caption{SKA predictions for the constraints on the growth of structures index (including BOSS and Planck priors) \cite{raccanelli}.}
\label{fig:gamma}
\end{figure}

As stated, testing the nature of dark matter or dark energy is a key issue in Cosmology. However, the effect of these putative entities can be manifestations of a yet unknown gravity theory beyond General Relativity \cite{whatif}. Thus, testing gravity at the strong limit provides a deeper understanding on the nature of this fundamental interaction. SKA, in particular, will be able to detect a vast number of pulsars, and binary systems of a pulsar orbiting a black hole, hence providing rich information about strong gravity. Thousands of millisecond pulsars might be detected, forming a “pulsar timing array” which can be a prime arena for detection of gravitational waves \cite{strong}.  On the other hand, General Relativity and alternative theories of gravity will be tested through gravitational lensing observations. In general, there are three model dependent parameters which characterise the growth of structures in any modified gravity theory \cite{euclid}. The first is the modified gravitational constant, $G_N \tilde{\mu}(a,k)$, where $\tilde{\mu}(a,k)$ is a model dependent correction in the Poisson equation, which can be expressed in the Fourier space as:
\begin{equation}
-2k^2\Phi = 8\pi G_N a^2\rho D \tilde{\mu} (a,k) ~,
\end{equation}
where $k$ is the wavenumber and $D$ is the gauge-invariant density contrast. The second is the anisotropic stress or slip relation, $\tilde{\gamma}$, which relates the two gauge invariant Bardeen potentials, $\Phi$ and $\Psi$, which come from time and spacial scalar perturbations of the metric, respectively:
\begin{equation}
\Psi = \tilde{\mu}(a,k)\Phi~,
\end{equation}
and the third one which is the growth rate, $f(a,k)$, (or its index $\gamma$):
\begin{equation}
f(a,k)=\left(\frac{a^2 8\pi G_N \rho}{3H^2}\right)^{\gamma}~,
\end{equation}
where $H=\dot{a}/a$ is the Hubble expansion rate. Therefore, this may discriminate between models of gravity. In Fig. \ref{fig:gamma}, we can see the improvement of the measurement of the equation of state parameter for dark energy over the index of structure growth. In General Relativity, $\tilde{\mu}=\tilde{\gamma}=1$  and  $\gamma\approx 0.545$. Thus, observational deviations from these values are evidences of modified gravity models, which can allow for predicting dependencies on the scale factor and on the wavenumber on the previous parameters. A particular example of such modified gravity theories is the one which admits an extension of the well known $f(R)$ theories with a non-minimal coupling between matter and curvature \cite{nmc}, whose cosmological perturbations give specific relationships for those parameters \cite{frazao}.

	Furthermore, it would also be interesting to investigate clusters rich in radiogalaxies in order to assess whether there is evidence for interacting dark matter-dark energy models \cite{ide1,ide2,ide3} or modified gravity theories \cite{lieq} through deviations from the virial theorem, which, at cosmic scales, is given by the so-called Layzer-Irvine equation.

\subsection{Cosmic magnetism}

Another relevant science objective of SKA is related to the cosmic magnetism, which is ubiquitous in the Universe, since interstellar gas, planets, stars and galaxies all exhibit magnetic fields. However, the shape and strength of such magnetic field in galaxies or even its origin is not known. The Universe itself can be magnetic. Therefore, radio observations can measure the Faraday rotation, the polarised synchrotron emission and the Zeeman effect, which gives a detailed information on such fields, in particular whether they are primordial\cite{turner,magnetic} or are generated later on via a dynamo mechanism. Another relevant issue is the impact of magnetic fields on the evolution of the Universe.

\section{CONCLUSIONS}
In this brief contribution we have presented the main contributions that SKA is expected to provide for Cosmology. As the world largest radiotelescope array, SKA will be able to access the parameters that characterise the dark components of the Universe with unprecedented detail: one expects to get a power spectrum for dark matter from weak lensing observations and to considerably constrain the equation of state of dark energy (cf. Eq. \ref{eq:1}). The acquired knowledge will allow for ruling out many models of dark matter and dark energy, including those where these two components interact with each other\cite{ide1,ide2,ide3} or come from a unique field or a fluid model\cite{kamenshchik,chaplygin, bilic}. Moreover, SKA data will provide essential information on putative alternative models of gravity beyond General Relativity.
	
	As mentioned, SKA will allow to look at the strong regime of Einstein’s gravity and of any other gravitational models. On the other hand, by studying weak lensing distribution of matter and gravitational distortions, SKA will provide means to test the three parameters that characterise alternative theories of gravity: the growth of structure factor (or index), the modified gravitational constant and the anisotropic stress.
	
	It is also expected that SKA will be an excellent tool to study cosmic magnetism and its characterisation over several redshifts, as well as the role of magnetic fields on the evolution of the Universe.
	
	To conclude we can say that SKA will, most likely, revolutionise the way we perceive the Universe, as it will provide a huge amount of data on a wide range of astrophysical and cosmological issues and will allow for the understanding of many key features of the Cosmos. 

%%%%%%%%%%%%%%%%%%%%%%%%%%%%%%%%%%%%%%%%%%%%%%%%%%%%%%%%%%%%%
\acknowledgments     %>>>> equivalent to \section*{ACKNOWLEDGMENTS}       
 
C. G. acknowledges the support from Fundação para a Ciência e a Tecnologia (FCT) under the grant SFRH/BD/102820/2014.   

%%%%%%%%%%%%%%%%%%%%%%%%%%%%%%%%%%%%%%%%%%%%%%%%%%%%%%%%%%%%%
%%%%% References %%%%%

%\bibliography{report}   %>>>> bibliography data in report.bib
%\bibliographystyle{spiebib}   %>>>> makes bibtex use spiebib.bst

\end{document}